\documentclass[a4paper,11pt]{article}
\usepackage{comment}

\usepackage{pos}
\usepackage{lipsum} 
\usepackage{graphicx}
\newlength{\bibitemsep}
\newlength{\bibparskip}
\let\oldthebibliography\thebibliography
\renewcommand\thebibliography[1]{%
  \oldthebibliography{#1}%
  \setlength{\parskip}{\bibitemsep}%
  \setlength{\itemsep}{\bibparskip}%
}

\usepackage{lineno}
\usepackage{babel}[english]

\title{Archival Search for IceCube Sub-TeV Neutrino Counterparts to Sub-Threshold Gravitational Wave Events from the Third Observing Run of the LIGO-Virgo-KAGRA}

\ShortTitle{IceCube Sub-TeV Neutrino Counterparts to Sub-Threshold GW from LVK O3}

\author{The IceCube Collaboration \\{\normalsize \normalfont(a complete list of authors can be found at the end of the proceedings)}\\}

\emailAdd{tista.mukherjee@kit.edu}

\abstract{

The IceCube Neutrino Observatory actively participates in multimessenger follow-ups of gravitational-wave (GW) events. With the release of the Gravitational-Wave Transient Catalogue (GWTC)-2.1 and -3, the sub-threshold GW event information from the third observation run of the LIGO-Virgo-KAGRA (LVK) detectors is publicly available.
These sub-threshold GWs are identified via template-based and minimally modelled search pipelines. Neutrino counterparts can enhance their astrophysical significance and improve their localisation. In this contribution, we propose a catalogue-based search for sub-TeV neutrino counterparts to sub-threshold GWs. For this search, we use archival data from IceCube’s dense infill array, DeepCore. Using the unbinned maximum likelihood method, we search for correlation between IceCube sub-TeV neutrinos and the $\sim$100 most significant sub-threshold GW source candidates. With this study, we aim to contribute to the ongoing efforts to identify common astrophysical sources of neutrinos and GWs. We present the current status of this search and its role in advancing multimessenger astronomy, paving the way for deeper exploration of GW events and their sources.

\vspace{4mm}

{\bfseries Corresponding authors:}
Tista Mukherjee$^{1*}$, 
\\
{$^{1}$ \itshape Institute for Astroparticle Physics, Karlsruhe Institute of Technology, Karlsruhe, Germany}\\
$^*$ Presenter
}

\ConferenceLogo{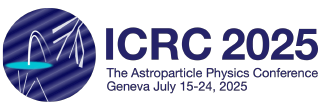}

\FullConference{39th International Cosmic Ray Conference (ICRC2025)\\
 15–24 July 2025\\
Geneva, Switzerland\\}

\begin{document}

\maketitle

\section{Introduction}
\label{intro}
The detection of gravitational waves (GWs) \cite{GW150914} and astrophysical neutrinos \cite{IC_AstroNu_2013} has established two of the foundational pillars of multimessenger astronomy. While joint GW-electromagnetic (EM) \cite{GW170817} and neutrino–EM detections \cite{TXS_IC} have been reported, no confirmed common source of GWs and high-energy neutrinos has yet been observed.

To search for such associations, the IceCube Neutrino Observatory 
has been actively following up on GW alerts from the LIGO–Virgo–KAGRA (LVK) collaboration, both via archival and real-time searches. By the end of O3, 90 of them were of `high-significance', typically with a false alarm rate (FAR) < 2 $\mathrm{yr^{-1}}$. They are most likely to originate from compact binary coalescence (CBCs). In addition to these events, the Gravitational Wave Transient Catalogues (GWTC)-2.1 \cite{gwtc-2.1} and -3 \cite{gwtc-3} also include a large set of lower-significance, or `sub-threshold' GW candidates from the third observing run (O3), which are now publicly available. These events are characterised by a more relaxed threshold of FAR < 2 $\mathrm{day^{-1}}$. As they were not followed up in real-time during O3, astrophysical origins for these events were not identified. Hence, they are interesting targets for archival neutrino searches. Identifying spatially and temporally coincident neutrino events can help elevate the significance of the sub-threshold GWs and improve their localisation \cite{elevated_pastro}.

A Bayesian framework was previously employed to assess the astrophysical relevance of these events based on high-energy neutrino data \cite{LLAMA}. However, the majority of the potential joint GW-neutrino sources are believed to have off-axis jets, leading to Doppler deboosting and consequently a dominant neutrino emission in the sub-TeV energy range \cite{Biehl_offAxisGRB}. Therefore, in this contribution, we present a complementary archival search for sub-TeV neutrino counterparts to the $\sim$100 most significant sub-threshold GW candidates from O3, as an extension of the analysis idea proposed in \cite{ICRC23_proceeding, ECRS24_TM}. The analysis uses data from IceCube’s DeepCore sub-array and employs an unbinned maximum likelihood (UML) method to assess spatial and temporal correlations. This work aims to enhance the astrophysical interpretation of sub-threshold GW events and contribute to the broader goal of identifying common sources of GWs and neutrinos.
 

\section{Datasets}
\label{sec2}

To carry out this analysis, we use publicly available sub-threshold GWs detected during O3 from the GWTC-2.1 and GWTC-3 data releases. For the neutrino component, we employed the sub-TeV neutrino data collected by IceCube-Deepcore. This section provides a brief overview of these datasets.

\subsection{Sub-threshold GWs from O3}
After the completion of the first half of O3 (O3a), a list of 1201 sub-threshold CBCs was included in GWTC-2.1. They were identified through four analysis pipelines tailored to detect CBCs such as binary black holes (BBH), binary neutron stars (BNS), and neutron star-black hole (NSBH) binaries. These CBC search pipelines are named GstLAL \cite{gstlal}, MBTA \cite{mbta}, PyCBC \cite{pycbc}, and PyCBC-Highmass \cite{pycbc-HM}. Similarly, 1048 sub-threshold GWs were further added to the GWTC-3 catalogue. This included not only CBCs from O3b, but also the CBC optimised version of `Coherent Wave Burst' (cWB) search pipeline \cite{cwb_allsky}, to identify GW burst-like events detected during the entire O3. The events obtained from this particular pipeline are referred to as 'cWBs' from here on in this proceeding.

\begin{figure}[htb!]
    \centering
    \includegraphics[width=0.5\linewidth]{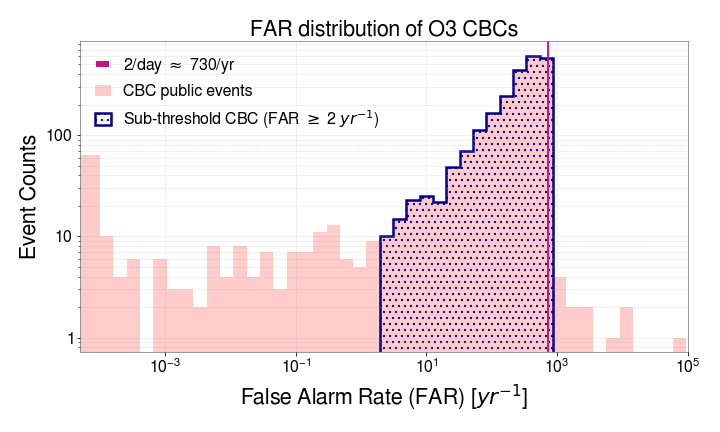}
    \includegraphics[width=0.49\linewidth]{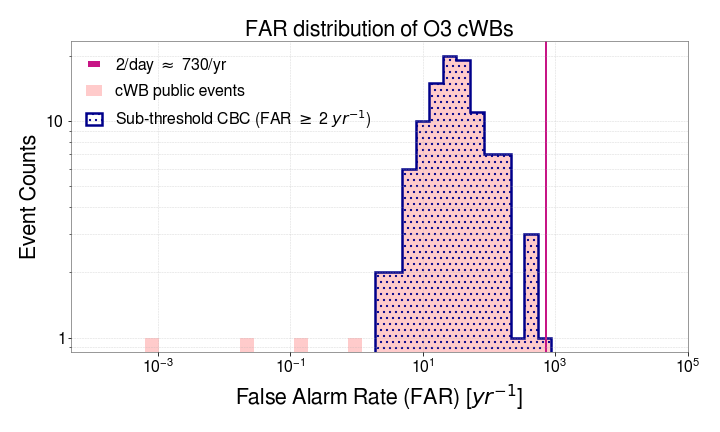}
    \caption{False Alarm Rate (FAR) distributions for CBCs (left) and cWBs (right) from O3. Four template-based pipelines identified CBC candidates during O3. Among them, those having FAR < 2 $\mathrm{yr^{-1}}$ are classified as confident detections. The cWB pipeline performs a minimally modelled search for burst-like GW events. In both panels, the dotted region marks sub-threshold events with $2~\mathrm{yr^{-1}} \leq \mathrm{FAR} < 2~\mathrm{day^{-1}}$, which are the focus of this study. Events above this range (right of the dashed line) are excluded. cWBs with FAR < $2~\mathrm{yr^{-1}}$, those not confirmed by CBC searches, remain scientifically interesting. Three of them were selected in this analysis for neutrino follow-up.
    }
    \label{fig:FAR}
\end{figure}

Unlike the current LVK selection criteria for high-significance detections, these sub-threshold GWs are expected to have $\mathrm{p_{astro}}$ < 0.5, alongside having FAR < 2 $\mathrm{day^{-1}}$. The quantity $\mathrm{p_{astro}}$ is the Bayesian probability of detecting a GW signal coming from a CBC-like source. Mathematically, we can express it as, $\mathrm{p_{astro}}$ = $\mathrm{p_{BBH} + p_{BNS} + p_{NSBH}}$.
Here, $\mathrm{p_{BBH}, ~p_{BNS}}$, and $\mathrm{p_{NSBH}}$ are the probabilities for detecting the GW signal from a BBH, BNS, and NSBH-like source, respectively. The FAR distribution for all CBCs and cWBs from O3 is shown in Fig. \ref{fig:FAR}.

As seen in the figure, GWTC-2.1 and GWTC-3 include all CBC and cWB candidates with FAR < 2 $\mathrm{day^{-1}}$ (approximately 730 $\mathrm{yr^{-1}}$). This threshold encompasses both low-significance and high-significance events.
Hence, as the first step, we exclude these high-significance GWs and focus on the sub-threshold candidates with $2~\mathrm{yr^{-1}} \leq \mathrm{FAR} < 2~\mathrm{day^{-1}}$. We follow the same selection procedure explained in \cite{ICRC23_proceeding}.

\begin{figure}[htb!]
    \centering
    \includegraphics[width=0.52\linewidth]{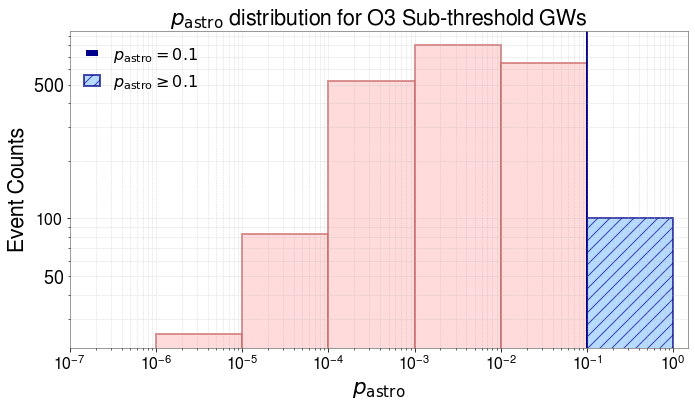}
    \includegraphics[width=0.46\linewidth]{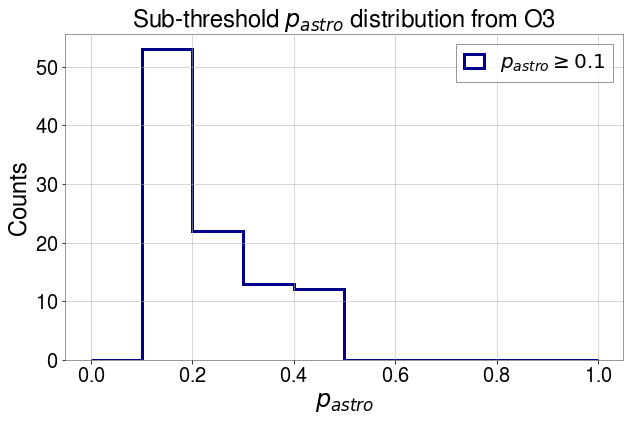}
    \caption{(Left) Distribution of $p_\mathrm{astro}$ values for all sub-threshold GWs with 2/yr < FAR < 2/day. It includes both CBCs and cWB triggers detected during O3. The last bin, highlighted in pale blue with hatching, includes events with $p_\mathrm{astro} \geq 0.1$. These are selected for the archival neutrino counterpart search presented in this work.
    (Right) Detailed $p_\mathrm{astro}$ distribution of the 100 shortlisted sub-threshold GWs within the range $0.1 \leq p_\mathrm{astro} \leq 1$. No events with $p_\mathrm{astro} \geq 0.5$ are present in this sample, consistent with the sub-threshold classification defined in \cite{gwtc-3}.}
    \label{fig:pastro}
\end{figure}

We then examine their $p_\mathrm{astro}$ distribution, shown in Fig. \ref{fig:pastro}. More than 90\% of these sub-threshold GWs were found to have $p_\mathrm{astro} \simeq 0$, indicating a high likelihood of being accidental triggers caused by terrestrial noise. Hence, we further refine our selection by only selecting events 0.1 $\leq p_\mathrm{astro} \leq$ 0.5. Finally, we have 100 shortlisted GWs for follow-up. Their $\mathrm{p_{astro}}$ distribution is shown in Fig. \ref{fig:pastro}.

\begin{figure}[htb!]
    \begin{minipage}{0.75\linewidth}
    \includegraphics[width=\linewidth]{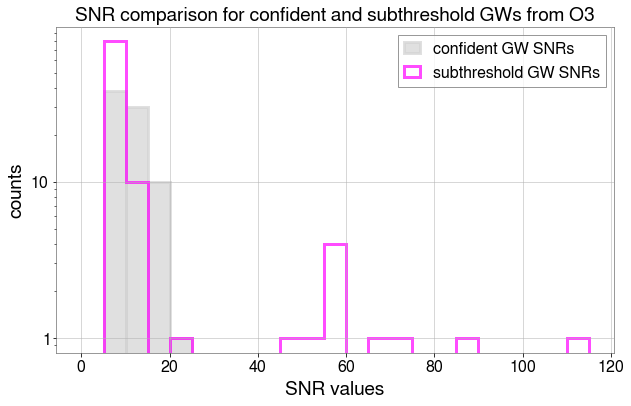}
    \end{minipage}
    \begin{minipage}{0.2\linewidth}
    \caption{The SNR distribution of the 100 shortlisted sub-threshold GWs from O3. A comparison is shown with the SNR distribution of the 79 confidently detected GWs during O3.}
    \label{fig:snr}
    \end{minipage}
\end{figure}

Notably, their corresponding signal-to-noise ratio (SNR) values are comparable to those of confidently detected GWs from O3, with a subset of events exhibiting particularly high SNRs, precisely SNR > 40, as presented in Fig. \ref{fig:snr}. This indicates that the selected candidates are not intrinsically weak signals. 
Rather, their classification as low-significance detections is likely due to their signal morphology deviating from typical CBC waveforms. 

In addition to the 100 shortlisted sub-threshold GWs, we identified 4 cWBs from Fig. \ref{fig:FAR} with FAR < 2 $\mathrm{yr^{-1}}$. They also were found to have $p_\mathrm{astro} > 0.5$ and exceptionally high SNRs (> 100). Eventually, three of these cWBs were found not to be associated with any confident CBC detection from O3. Due to their strong signal characteristics, they are included in our follow-up analysis.
Hence, in total, our search targets 103 sub-threshold GW candidates for possible sub-TeV neutrino counterparts using IceCube data.

\subsection{IceCube sub-TeV neutrinos}
The sub-TeV neutrino dataset available within IceCube is called `GRECO' (GeV Reconstructed Events with Containment for Oscillations). This is an all-sky and all-flavour dataset suitable for transient follow-up searches
It covers the energy range of O(10-100) GeV.
It is characterised by a highly stable event rate ($\sim$ 5 mHz) and good effective area coverage across this energy range, as shown in \cite{novae}. This makes it suitable for transient searches.
Previous studies (e.g., \cite{greco-gw}) have used GRECO to investigate possible lower-energy neutrino counterparts to 90 confident GW events observed during the O1, O2, and O3 runs. Here, we extend this approach by using GRECO to search for sub-TeV neutrino counterparts associated with shortlisted sub-threshold GW candidates from the O3 run.

\section{Analysis Method}
\label{sec3}
To search for the neutrino counterparts, we will follow an Unbinned Maximum Likelihood (UML) analysis. We define the following likelihood function:
\begin{equation}
\label{likelihood}
    \mathcal{L} (n_\mathrm{s} (\gamma)) = \frac{(n_\mathrm{s} + n_\mathrm{b})^N}{N!} \mathrm e^{-(n_\mathrm{s} + n_\mathrm{b})} \prod_{i = 1}^{N} \Big ( \frac{n_\mathrm{s} S_\mathrm{i}}{n_\mathrm{s} + n_\mathrm{b}} + \frac{n_\mathrm{b} B_\mathrm{i}}{n_\mathrm{s} + n_\mathrm{b}} \Big ).
\end{equation}

Here, N is the total number of events observed in the sky. Among those N events, $n_\mathrm{s}$ is the number of signal neutrinos. So, the expected number of background events, $n_\mathrm{b}$ = N - $n_\mathrm{s}$.
The index $i$ runs over each of the $N$ candidate neutrino events. The signal and background probability density function (PDF) for the $i^\mathrm{{th}}$ event, denoted as $S_\mathrm{i}$ and $B_\mathrm{i}$, are defined following the method explained in \cite{greco-gw}.

Using this likelihood function, we define a test statistic (TS).
\begin{equation}
\label{TS}
    TS = 
    \Bigg[ 2~\mathrm{ln}  \Bigg( \frac{\mathcal{L}_\mathrm{k} (n_\mathrm{s}(\gamma)) \cdot \omega_\mathrm{k}}{\mathcal{L}_\mathrm{k} (n_\mathrm{s} = 0) } \Bigg) \Bigg ].
\end{equation}

Here, $k$ is the index of each pixel in the sky, and $\omega_{k}$ is a spatial prior term at that location. For each pixel, $\omega_\mathrm{k}$ is scaled linearly with the probability of having a GW source in that pixel. The details of each of these components and the entire framework for analysis with sub-threshold GW candidates have been described in \cite{ICRC23_proceeding}.

\section{Results}
\label{sec4}

\begin{figure}[htb!]
    \centering
    \includegraphics[width=0.75\linewidth]{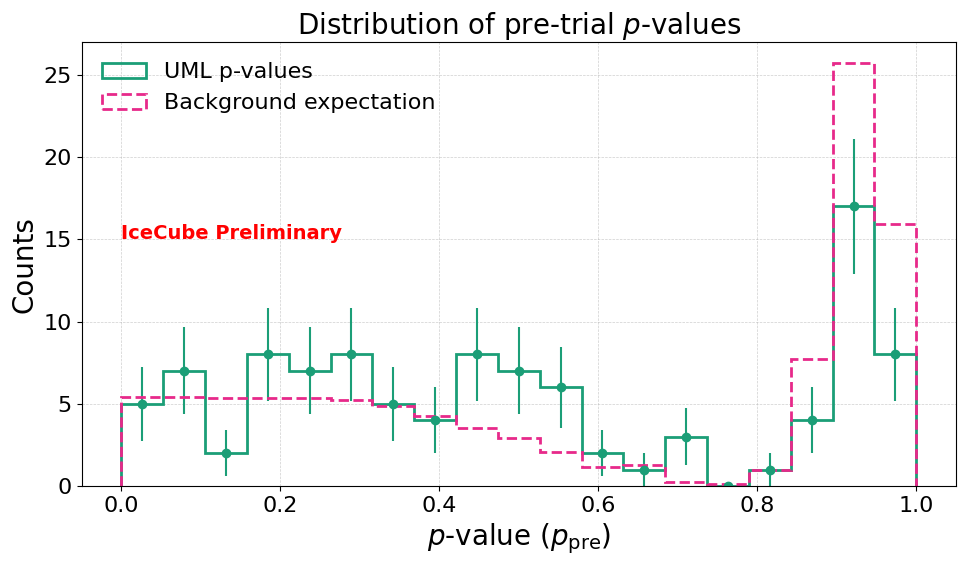}
    \caption{Distribution of pre-trial $p$-values for the 103 shortlisted sub-threshold GW events from the O3 observing run. The green histogram with error bars shows the observed $p$-value distribution following the UML method. It is compared with the expected background distribution, obtained by randomly sampling TS values from the background TS distribution for each GW event and calculating the corresponding $p$-values. This background sampling was repeated 10,000 times to construct the expectation. The observed distribution is consistent with the null hypothesis, and no significant excess is observed.}
    \label{fig:pval}
\end{figure}

For the 103 shortlisted GW events, we conduct a catalogue search to identify spatial and temporal coincidences with the events from the GRECO dataset within a $\pm$ 500 s time window around the GW event detection time. The results of this analysis are summarised here. The distribution of pre-trial $p$-values is shown in Fig. \ref{fig:pval}. It is compared with the expected background distribution. That is obtained by randomly sampling TS values from the background TS distribution for each GW event and calculating the corresponding $p$-values. This background sampling was repeated 10,000 times to construct the expectation. As the observed $p$-values are compared with the corresponding background expectation, no significant deviation from the null hypothesis is apparent. Of the selected events, 10 were sub-threshold cWBs, 16 are NSBH-like, and 3 are cWBs with $p_\mathrm{astro} > 0.5$. None of these show statistically significant neutrino correlations within the search time window.

\begin{figure}[htb!]
    \centering
    \includegraphics[width=0.55\linewidth]{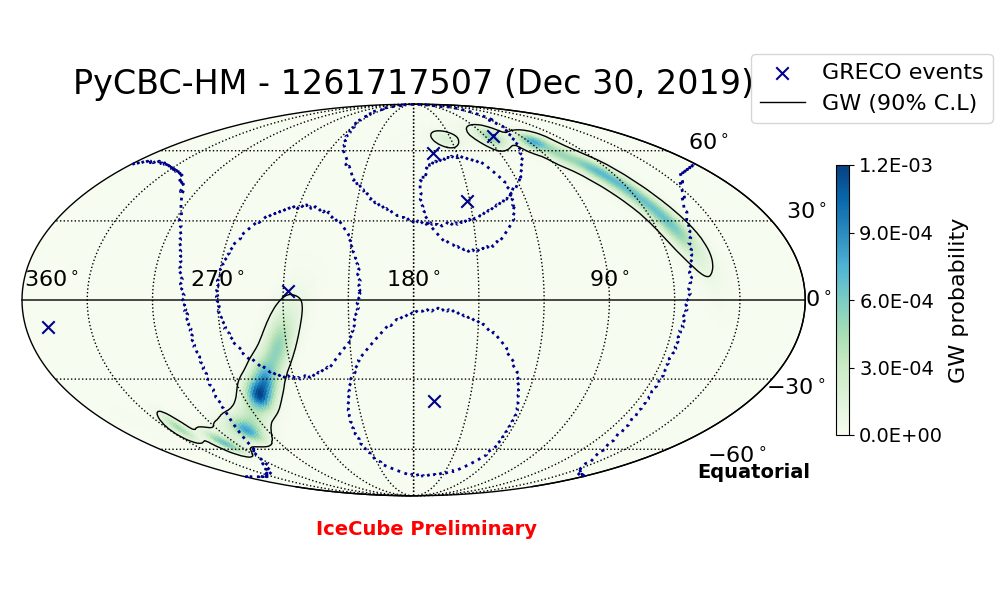}
    \includegraphics[width=0.44\linewidth]
    {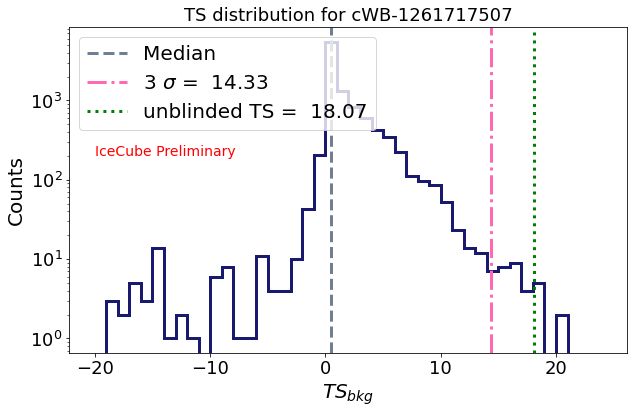}
    \caption{Skymap of the most significant sub-threshold GW from the shortlisted 103 events. The GPS time for detection of the GW event is 1261717507, which corresponds to December 30, 2019. The $p_{\mathrm{astro}}$ value corresponding to this detection is 0.16. The neutrinos detected within 1000 s around the sub-threshold GW detection time are overlaid on top of the GW skymap. The probability of chance coincidence for this event is 0.038\% (4\% after trial correction).
    (Right) The observed TS value is shown here, overlaid on the corresponding background TS distribution. The median of the background distribution and 3$\sigma$ deviation from it are also indicated.}
    \label{fig:CBC_skymap}
\end{figure}

The most significant candidate is a sub-threshold BBH-like event. The GPS time of the event is 1261717507. It was detected by the PyCBC-Highmass pipeline on December 30, 2019, during O3b. The pre-trial $p$-value associated with this event is 0.00038 (3.35 $\sigma$). However, after trial-correction, the significance drops to 0.04 (1.76 $\sigma$). The joint skymap for the GW and GRECO events detected within 1000 s by IceCube is shown in Fig. \ref{fig:CBC_skymap}.

\begin{figure}[htb!]
    \centering
    \includegraphics[width=0.55\linewidth]{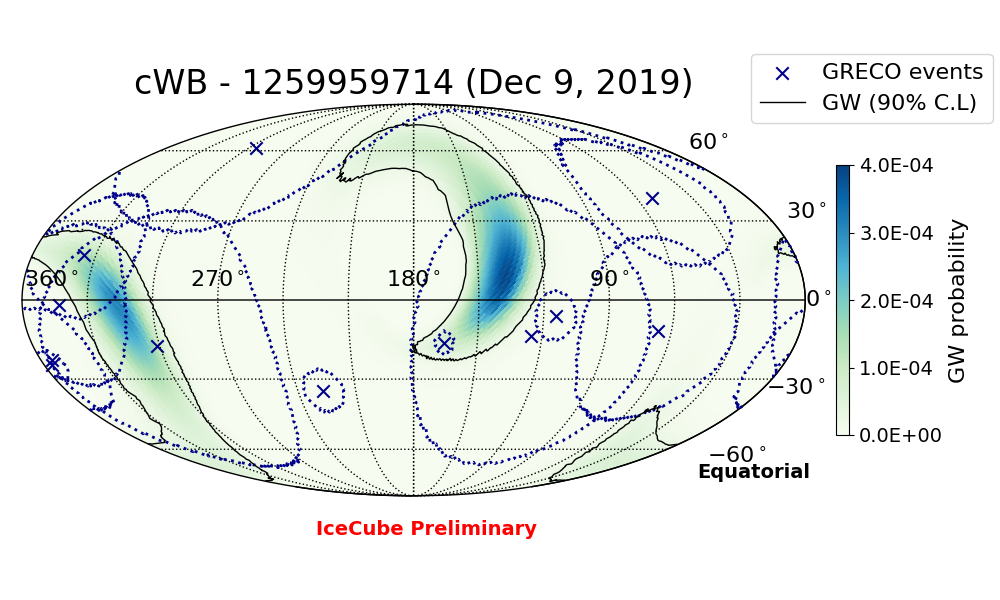}
    \includegraphics[width=0.44\linewidth]{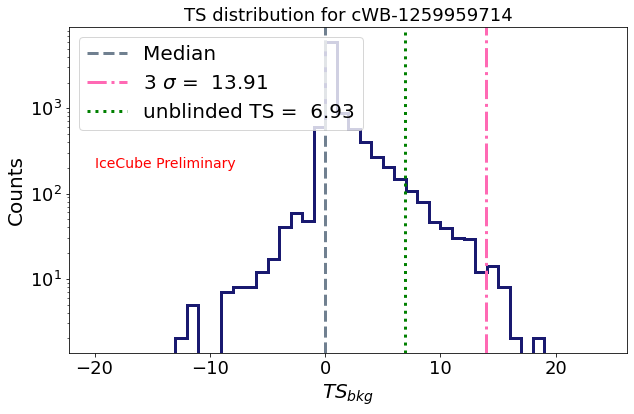}
    \caption{(Left) Skymap of the most significant sub-threshold cWB from the shortlisted 103 events. The GPS time for detection of the GW event is 1259959714, which corresponds to December 9, 2019. The $p_{\mathrm{astro}}$ value for this detection is 0.28. Neutrinos detected within the 1000~s search window are overlaid on the GW skymap. The pre-trial $p$-value for this event is 0.038. (Right) The observed TS value is overlaid on the corresponding background TS distribution. The median and the $3\sigma$ range of the background are also indicated.}
    \label{fig:cWB_skymap}
\end{figure}

Among the shortlisted cWBs, the most significant one has a GPS time of 1259959714. It was detected again during O3b, on December 9, 2019, solely by the CBC optimised GW burst search pipeline. The pre-trial $p$-value was found to be 0.038.
The joint GW skymap with GRECO events detected within 1000 s by IceCube is shown in Fig. \ref{fig:cWB_skymap}. 

The right panels of Fig. \ref{fig:CBC_skymap} and \ref{fig:cWB_skymap} show the observed TS values for the most significant sub-threshold CBC and cWB events, overlaid on the respective background TS distributions derived from scrambled trials. The median and the $3\sigma$ deviation from the background are also marked for reference.

As none of the shortlisted events has > 3 $\sigma$ significance after trial correction, we report flux upper limits (ULs) (90\% C.L.) for all of them. These limits were calculated assuming an $E^{-2}$ neutrino spectrum and a fixed 1000 s integration time around each GW detection time. The flux ULs are plotted as a function of the best-fit declination obtained solely from the corresponding GW skymaps. For context, we also include previously published ULs for 79 confident GW events from O3, obtained using both the GRECO and GFU datasets \cite{greco-gw, GWfollowupO3}. The comparison shows that the ULs with GRECO for sub-threshold GWs are of a similar order of magnitude to those for confident events. The limits obtained with GFU are more constraining, as expected, primarily due to improved angular resolution for high-energy tracks. It allows better suppression of background within a smaller search region. This enhances the signal-to-background separation and leads to better flux ULs.


\begin{figure}
    \begin{minipage}{0.7\linewidth}
    \includegraphics[width=\linewidth]{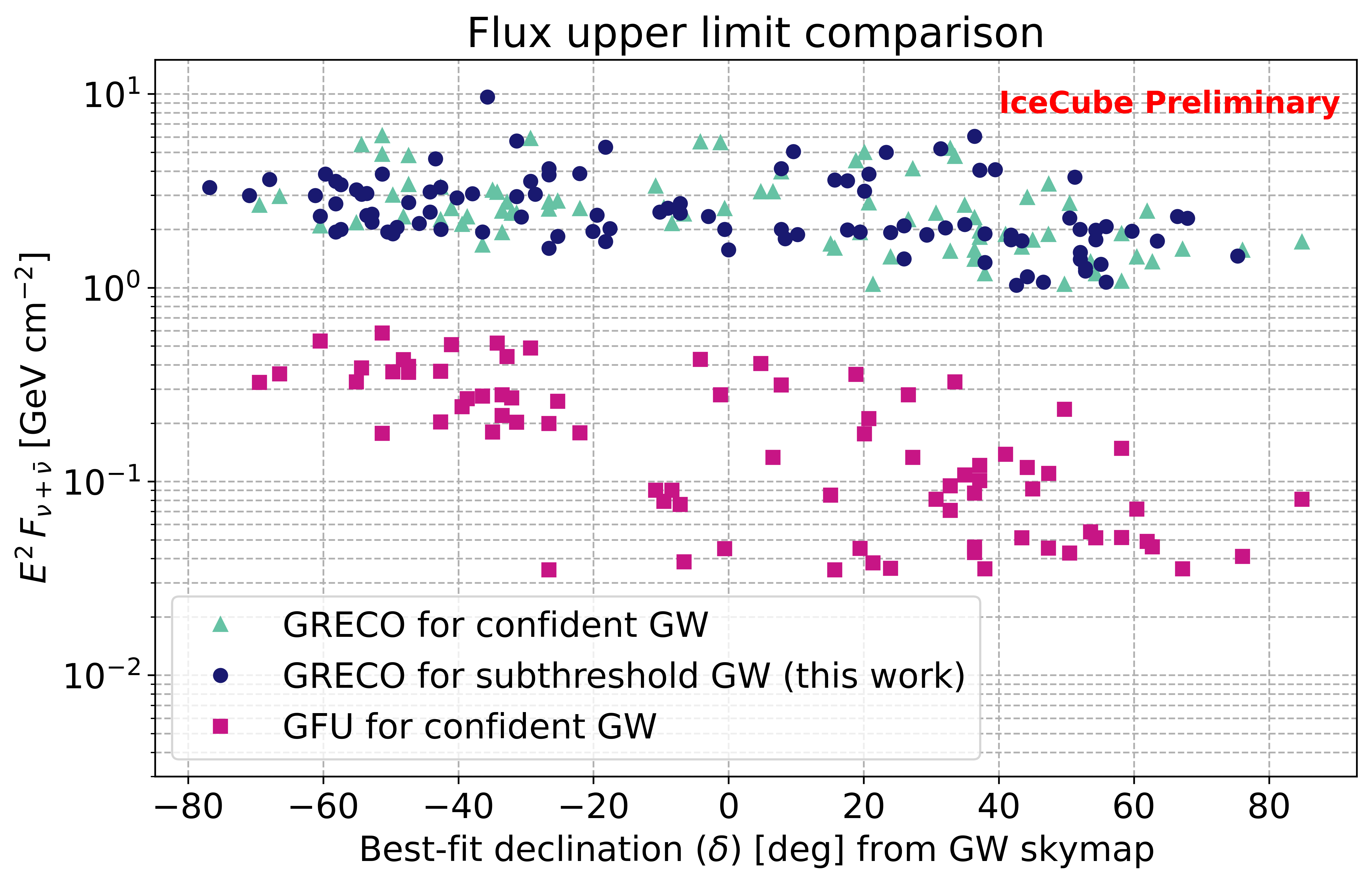}
    \end{minipage}
    \begin{minipage}{0.29\linewidth}
    \caption{90\% C.L. flux upper limits for neutrino emission associated with the 103 shortlisted sub-threshold GW events. The calculation is performed assuming spectral index $\gamma$ = 2 and a 1000-second integration time. Results are compared to previously published ULs for the 79 confident GW events from O3, using both GRECO \cite{greco-gw} and GFU datasets \cite{GWfollowupO3}. Each flux value is plotted against the best-fit declination from the corresponding GW skymap.}
    \label{fig:flux_UL}
    \end{minipage}
\end{figure}

\section{Summary}
As sub-threshold GWs represent valuable targets for multimessenger searches, in this contribution, we present an archival UML follow-up search for sub-TeV neutrinos in temporal and spatial coincidence with sub-threshold GWs from O3. No statistically significant neutrino counterpart was identified after accounting for trial factors. Consequently, neutrino flux ULs were derived for each event. The analysis framework employed here can be readily extended to other IceCube data samples. With sub-threshold GW alerts now publicly available in real time during O4, this approach opens promising avenues for future real-time multimessenger campaigns.


\bibliographystyle{ICRC}
\bibliography{references}

\clearpage

\section*{Full Author List: IceCube Collaboration}

\scriptsize
\noindent
R. Abbasi$^{16}$,
M. Ackermann$^{63}$,
J. Adams$^{17}$,
S. K. Agarwalla$^{39,\: {\rm a}}$,
J. A. Aguilar$^{10}$,
M. Ahlers$^{21}$,
J.M. Alameddine$^{22}$,
S. Ali$^{35}$,
N. M. Amin$^{43}$,
K. Andeen$^{41}$
% ...
\\
\\
% ...
$^{10}$ Universit{\'e} Libre de Bruxelles, Science Faculty CP230, B-1050 Brussels, Belgium \\
% ...
$^{16}$ Department of Physics, Loyola University Chicago, Chicago, IL 60660, USA \\
$^{17}$ Dept. of Physics and Astronomy, University of Canterbury, Private Bag 4800, Christchurch, New Zealand \\
% ...
$^{21}$ Niels Bohr Institute, University of Copenhagen, DK-2100 Copenhagen, Denmark \\
$^{22}$ Dept. of Physics, TU Dortmund University, D-44221 Dortmund, Germany \\
% ...
$^{35}$ Dept. of Physics and Astronomy, University of Kansas, Lawrence, KS 66045, USA \\
% ...
$^{39}$ Dept. of Physics and Wisconsin IceCube Particle Astrophysics Center, University of Wisconsin{\textemdash}Madison, Madison, WI 53706, USA \\
% ...
$^{41}$ Department of Physics, Marquette University, Milwaukee, WI 53201, USA \\
% ...
$^{43}$ Bartol Research Institute and Dept. of Physics and Astronomy, University of Delaware, Newark, DE 19716, USA \\
% ...
$^{63}$ Deutsches Elektronen-Synchrotron DESY, Platanenallee 6, D-15738 Zeuthen, Germany \\
$^{\rm a}$ also at Institute of Physics, Sachivalaya Marg, Sainik School Post, Bhubaneswar 751005, India \\
% ...

\subsection*{Acknowledgments}

\noindent
The authors gratefully acknowledge the support from the following agencies and institutions:
USA {\textendash} U.S. National Science Foundation-Office of Polar Programs,
U.S. National Science Foundation-Physics Division,
U.S. National Science Foundation-EPSCoR,
U.S. National Science Foundation-Office of Advanced Cyberinfrastructure,
Wisconsin Alumni Research Foundation,
Center for High Throughput Computing (CHTC) at the University of Wisconsin{\textendash}Madison,
Open Science Grid (OSG),
Partnership to Advance Throughput Computing (PATh),
Advanced Cyberinfrastructure Coordination Ecosystem: Services {\&} Support (ACCESS),
Frontera and Ranch computing project at the Texas Advanced Computing Center,
U.S. Department of Energy-National Energy Research Scientific Computing Center,
Particle astrophysics research computing center at the University of Maryland,
Institute for Cyber-Enabled Research at Michigan State University,
Astroparticle physics computational facility at Marquette University,
NVIDIA Corporation,
and Google Cloud Platform;
Belgium {\textendash} Funds for Scientific Research (FRS-FNRS and FWO),
FWO Odysseus and Big Science programmes,
and Belgian Federal Science Policy Office (Belspo);
Germany {\textendash} Bundesministerium f{\"u}r Forschung, Technologie und Raumfahrt (BMFTR),
Deutsche Forschungsgemeinschaft (DFG),
Helmholtz Alliance for Astroparticle Physics (HAP),
Initiative and Networking Fund of the Helmholtz Association,
Deutsches Elektronen Synchrotron (DESY),
and High Performance Computing cluster of the RWTH Aachen;
Sweden {\textendash} Swedish Research Council,
Swedish Polar Research Secretariat,
Swedish National Infrastructure for Computing (SNIC),
and Knut and Alice Wallenberg Foundation;
European Union {\textendash} EGI Advanced Computing for research;
Australia {\textendash} Australian Research Council;
Canada {\textendash} Natural Sciences and Engineering Research Council of Canada,
Calcul Qu{\'e}bec, Compute Ontario, Canada Foundation for Innovation, WestGrid, and Digital Research Alliance of Canada;
Denmark {\textendash} Villum Fonden, Carlsberg Foundation, and European Commission;
New Zealand {\textendash} Marsden Fund;
Japan {\textendash} Japan Society for Promotion of Science (JSPS)
and Institute for Global Prominent Research (IGPR) of Chiba University;
Korea {\textendash} National Research Foundation of Korea (NRF);
Switzerland {\textendash} Swiss National Science Foundation (SNSF).

\end{document}